\documentclass[]{aastex631}

\usepackage{lipsum}
\usepackage{amsmath}
\usepackage{mathtools}
\usepackage[nottoc]{tocbibind}
\usepackage{verbatim}
\usepackage{soul}

\begin{document}

\title{Predicting High-Energy neutrino emission from X-ray absorption in CSM-interacting SNe}

\correspondingauthor{silvia.gagliardini@uniroma1.it}

\author[0000-0002-1832-2126]{Silvia Gagliardini}
\affil{Physics Department,Ariel University, Ariel, Israel}
\affil{Istituto Nazionale di Fisica Nucleare, Sezione di Roma, P. le Aldo Moro 2, I-00185 Rome, Italy}
\affil{Dipartimento di Fisica dell'Universit\`a La Sapienza, P. le Aldo Moro 2, I-00185 Rome, Italy}
\author[0000-0003-4366-8265]{Simone Dall'Osso}
\affil{INAF - Istituto di Radioastronomia, via Gobetti 101, Bologna, 40129, Italy}
%\affil{Dipartimento di Fisica e Astronomia -- Universit\`{a} di Bologna - Bologna, Italy}
\author[0000-0002-7349-1109]{Dafne Guetta}
\affil{Physics Department,Ariel University, Ariel, Israel}

\begin{abstract}

Core-collapse supernovae (CCSNe) interacting with a dense circumstellar medium (CSM) are promising sources of high-energy neutrinos. Most neutrino predictions rely on specific assumptions about the CSM density profile and progenitor mass-loss history. We develop a data-driven framework that connects X-ray observations directly to the expected neutrino signal.~Using the temporal evolution of the hydrogen-equivalent absorbing column density, $N_{\rm H}(t)$, together with the shock velocity inferred from the X-ray plasma temperature, we reconstruct the CSM density profile and shock velocity evolution. We estimate the shock-breakout (SBO) epoch and the resulting neutrino emission, and apply the method to the nearby interacting supernovae SN~2023ixf and SN~2024ggi.
For both sources, we compute the expected neutrino event rates for IceCube and KM3NeT/ARCA. We find that neutrino production is concentrated near SBO, when hadronic interactions are most efficient.~This naturally defines a short time window, specific to each source, that maximizes the signal-to-background ratio. Our results demonstrate that X-ray observations can constrain {\it at once} both the expected neutrino signal and the optimal time window for its detection, providing an observationally-driven strategy for multimessenger searches of these shock-powered transients.

\end{abstract}

\section{Introduction}

Core‑collapse supernovae (CCSNe) exploding into a dense circumstellar medium (CSM) have long been recognized as promising sites for high‑energy neutrino production (\citealt{Murase_2019,Waxman_2024}). When the rapidly expanding ejecta collide with the surrounding material, they drive a strong forward shock capable of accelerating cosmic rays (CRs) to high  energies (\citealt{Murase_2019, Gagliardini2026}). In environments where the CSM density is sufficiently high, such as in strongly interacting Type IIn and other CSM‑rich events, inelastic proton-proton ($pp$) interactions become efficient. These collisions produce charged pions that subsequently decay into high‑energy neutrinos. Recent observationally motivated studies, including analyses of SN2023ixf \citep{Guetta2023,Murase_2024,Cosentino2025}and other nearby interacting CCSNe, further support the idea that dense‑CSM shocks can serve as transient but powerful neutrino accelerators.

In light of these considerations, the diffuse astrophysical neutrino flux detected by IceCube remains a central unresolved problem in high‑energy astrophysics. The shock‑driven hadronic processes outlined above naturally position CCSNe embedded in dense CSM as a viable contributor to this flux. 
In dense circumstellar environments, efficient CR acceleration at the forward shock combined with a high target density naturally produces substantial neutrino emission through hadronic $pp$ interactions. A significant contribution to the diffuse neutrino background does not require the accompanying $\gamma$-rays to be suppressed. However, when the CSM is optically thick to high‑energy photons, the resulting attenuation of $\gamma$‑ray emission makes neutrinos the sole observable tracer of the underlying $pp$ interactions, and therefore of the particle acceleration processes and shock physics that drive them. In such cases, interacting CCSNe become uniquely valuable laboratories: they can both contribute meaningfully to the neutrino background and reveal particle acceleration in regimes where electromagnetic signatures are largely hidden.

The efficiency of particle acceleration in these environments depends critically on the physical nature of the shock. At very early times, the shock is expected to be radiation mediated (RMS), with radiation pressure dominating the shock structure and suppressing efficient diffusive shock acceleration. As the shock propagates outward and the optical depth of the surrounding medium decreases, it eventually transitions into a collisionless shock (CLS), enabling efficient acceleration of relativistic particles. 
In circumstellar environments with sufficiently high density, the RMS launched by the explosion does not break out at the stellar surface but instead propagates into the surrounding material. As the shock advances through the optically thick CSM, the local optical depth eventually drops enough to allow photons to escape, causing the shock breakout (SBO) to occur within the CSM itself. This CSM‑embedded breakout produces a bright X‑ray flash, a hallmark of radiation escaping from a dense environment. The emergence of this X‑ray emission signals a fundamental change in the shock regime: once photons can escape, radiation can no longer mediate the shock structure, and the shock transitions to a collisionless one capable of accelerating CRs efficiently \citep{Ofek_2013,Wasserman_2025}. Because efficient neutrino production requires both accelerated particles and a dense target for hadronic $pp$ interactions, this transition, when a CLS forms within a high‑density CSM and X‑rays begin to escape naturally marks the onset of high‑energy neutrino production.
In this work, we follow the methodology introduced by \cite{WaxmanLoeb2001} and employ the SBO condition as an observationally grounded indicator of the onset of efficient particle acceleration and hadronic interactions, i.e. of neutrino emission.

A key diagnostic of the circumstellar environment is the hydrogen‑equivalent column density, $N_H$, which can be inferred from photoelectric absorption features in the X‑ray spectrum. In dense CSM configurations, soft X‑rays are strongly attenuated by photoelectric absorption, while higher‑energy photons may still escape, allowing the spectral shape to probe the amount of intervening material. The temporal evolution of $N_H$ therefore provides a direct observational proxy of the CSM mass distribution surrounding the explosion. As the shock propagates outward and the absorbing column diminishes, the CSM becomes progressively transparent to X‑rays, revealing the underlying high‑energy emission.

Several recent studies \citep{Murase_2019,Murase_2024,Waxman_2024,Cosentino2025,Wasserman_2026} have explored high‑energy neutrino production in interacting supernovae using semi‑analytical parameterizations for the CSM density profile and shock evolution. Although these models have provided valuable insight into the conditions under which CCSNe may produce detectable neutrino emission, their predictions necessarily depend on assumed CSM structures that are often poorly constrained observationally. The increasing availability of high‑cadence X‑ray observations now offers a more direct avenue for characterizing the CSM environment and its evolution.

In interacting CCSNe, X‑ray emission typically emerges only a few days after explosion, indicating that the earliest phases of shock interaction occur within a highly opaque, dense CSM. The observed X‑ray spectrum, particularly the evolution of the absorbing column density and shock temperature, encodes the amount and distribution of CSM surrounding the progenitor. Consequently, the time‑dependent hydrogen‑equivalent column density, $N_H$, provides a direct, data‑driven probe of the CSM density profile and the shock’s dynamical evolution.

Recent semi‑analytical work, including applications to SN~2023ixf \citep{Cosentino2025}, has examined neutrino production in hydrogen‑rich CCSNe interacting with low‑mass CSM.~However, these studies rely on parameterized CSM descriptions that are not uniquely constrained by observations.~In contrast, our approach uses the observed evolution of the X‑ray absorbing column density and shock temperature to reconstruct the CSM structure and shock dynamics directly from the data. This enables us to derive the expected neutrino flux with significantly reduced dependence on assumed CSM properties, providing a more observationally-anchored prediction of neutrino emission from nearby interacting supernovae.

SN~2023ixf and SN~2024ggi provide excellent laboratories to investigate particle acceleration and high-energy neutrino production in interacting supernovae owing to their proximity and the availability of extensive early time X-ray observations. In particular, both events exhibit evidence for strong interaction between the supernova ejecta and a dense circumstellar environment, allowing the temporal evolution of the absorbing column density and shock properties to be directly constrained from observations. SN~2023ixf constitutes the primary case study of this work due to the exceptional quality of its X-ray data, which provides measurements of both the absorbing column density and the shock temperature with associated uncertainties over multiple epochs. These observations allow a fully data-driven reconstruction of the shock evolution, CSM density profile, SBO epoch, and neutrino emission, including uncertainty estimates.

SN~2024ggi is instead used as an independent test case of the proposed framework. Although extensive X-ray observations are available, the data quality is not as good as in SN~2023ixf, and the published temperature estimates prevent a complete evaluation of measurement errors on the shock velocity evolution, and on derived quantities. Nevertheless, SN~2024ggi offers an important opportunity to assess the applicability of the method to a second nearby interacting supernova with different CSM properties.

The paper is organized as follows. In Section~\ref{sec:2}, we introduce the physical connection between CSM interaction, SBO. In Section~\ref{sec:3}, we develop the data-driven framework that uses the observed X-ray absorption and plasma temperature to reconstruct the CSM density profile and shock evolution, determine the SBO epoch, and predict the resulting neutrino emission. In Section~\ref{sec:4}, we
describe the neutrino spectrum and detector response, and introduce the statistical framework used to assess detectability and determine the
optimal neutrino search window. We then present and discuss the results for SN~2023ixf and SN~2024ggi in Sections~\ref{sec:optimal_window}. Finally, in Section~\ref{sec:5}, we discuss the implications and systematic uncertainties of our results and summarize our conclusions.

\section{Shock interaction, CSM column density, shock breakout, and the RMS-CLS transition}
\label{sec:2}

The interaction between the expanding supernova ejecta and the surrounding CSM dissipates kinetic energy at the forward shock, giving rise to a shock‑powered luminosity, whose instantaneous shock value can be approximated as

\begin{equation}
L_{\rm sh}(t) \simeq
2\pi R_{\rm sh}^2(t)\,
\rho_{\rm csm}\!\left[R_{\rm sh}(t)\right]\,
v_{\rm sh}^3(t),
\label{eq:L_sh}
\end{equation}

where $R_{\rm sh}(t)$ and $v_{\rm sh}(t)$ denote the shock radius and velocity, respectively, and $\rho_{\rm csm}\!\left[R_{\rm sh}(t)\right]$ is the CSM density profile, $\rho_{\rm csm}(r)$, evaluated at the shock position, $r=R_{\rm sh}(t)$.

The kinetic energy dissipated at the forward shock is divided between thermal heating of the shocked gas, which powers the observed X‑ray luminosity, and non‑thermal channels such as the acceleration of relativistic protons that subsequently drive hadronic interactions. We parameterize this energy partition as $L_X = \epsilon_X L_{\rm sh}$ and $L_{\rm CR} = \epsilon_{\rm CR} L_{\rm sh}$ where $\epsilon_X$ and $\epsilon_{\rm CR}$ denote the fractions of the shock power converted into X-rays and accelerated CRs. The latter will undergo hadronic interactions leading to neutrino emission, so we can parametrize the neutrino luminosity as $
L_\nu \simeq f_{pp}\,\epsilon_{\rm CR}\,L_{\rm sh},$
where $f_{pp}$ represents the fraction of protons that undergo $pp$ interactions in the shocked CSM.

To connect the CSM density to the observable X-ray absorption, we introduce the hydrogen number density

\begin{equation}
n_H(r)=
\frac{\rho_{\rm csm}(r)}
{\mu_e m_p},
\end{equation}

where $\rho_{\rm csm}(r)$ is the CSM mass density, $m_p$ is the proton mass, and $\mu_e \simeq 1.2$ is the mean molecular weight per electron for a fully ionized plasma of approximately solar composition \citep{Draine2011}.

The hydrogen column density ahead of the shock is
\begin{equation}
N_H(R_{\rm sh})=
\int_{R_{\rm sh}}^\infty n_H(R)\,dR
\simeq n_H(R_{\rm sh})\,R_{\rm sh},
\label{eq:NH_R}
\end{equation}

showing that the observed hydrogen-equivalent column density directly traces the local density at the shock radius.

The Thomson optical depth ahead of the shock is

\begin{equation}
\tau_T(R_{\rm sh})=
\int_{R_{\rm sh}}^\infty
n_e(r) \, \sigma_T \, dr,
\end{equation}

where $n_e(r) =n_H(r)$ is the electron number density and $\sigma_T$ is the Thomson cross section.

The optical depth can therefore be expressed in terms of the hydrogen-equivalent column density as

\begin{equation}
\tau_T(R_{\rm sh})
\simeq
\mu_e \sigma_T N_H(R_{\rm sh}).
\label{eq:tau_T_R}
\end{equation}

SBO occurs when the optical depth ahead of the shock satisfies \citep{Waxman_2024}

\begin{equation}
\tau_T(t_{\rm bo})=\tau_{T,\rm bo}
\simeq
\frac{c}{v_{\rm sh}(t_{\rm bo})},
\label{eq:tau_T_t}
\end{equation}

where $v_{\rm sh}(t_{\rm bo})$ is the shock velocity at the breakout epoch ($t_{\rm bo}$). This condition marks the stage at which radiation can diffuse ahead of the shock and escape from the optically thick CSM. In the standard picture, this becomes associated with the transition from a RMS to a CLS, which enables efficient particle acceleration and, thus, hadronic interactions \citep{WaxmanLoeb2001}.
Using the relation between optical depth and column density, the breakout condition can be expressed in terms of a critical hydrogen-equivalent column density,

\begin{equation}
N_H(t_{\rm bo})
= \frac{c}
{\mu_e \sigma_T v_{\rm sh}(t_{\rm bo})}.
\label{eq:NH_bo}
\end{equation}

This relation provides a direct connection between the observed X-ray absorption and the onset of shock breakout. If the progenitor is surrounded by an extended dense CSM, SBO may occur well outside the stellar surface. In this scenario, radiation remains trapped until the shock has propagated through a sufficiently large fraction of the optically thick CSM. As a consequence, the emergence of observable X-ray emission can be delayed by several days with respect to the explosion epoch.

\section{A data-driven framework from X-ray absorption to neutrino emission}
\label{sec:3}

In this section we develop the framework that connects the observed X-ray absorption in interacting supernovae to the expected high-energy neutrino.
After a brief summary of the standard wind-like description of the CSM with constant shock velocity, we develop our data-driven approach in which neither the CSM density nor the shock velocity are fixed {\it a priori}.~Rather, they are both reconstructed from the measured plasma temperature and the observed temporal evolution of the hydrogen-equivalent column density, $N_H(t)$, as inferred from X-ray spectroscopy.

This approach allows us to estimate the shock power, the efficiency of hadronic interactions and the neutrino luminosity without assuming a particular progenitor mass-loss history or a steady-wind density profile.

\subsection{Standard wind-like description with constant shock velocity}

The CSM surrounding red supergiant (RSG) progenitors is commonly described as a stationary stellar wind with density

\begin{equation}
\rho_{\rm csm}(r)
=\frac{\dot{M}}
{4\pi r^{2} v_w},
\end{equation}

where $\dot{M}$ is the mass-loss rate of the progenitor and $v_w$ is the constant wind velocity. 

For this wind density profile, $\rho_{\rm csm}\propto r^{-2}$, the hydrogen-equivalent column density scales as $N_H(R_{\rm sh}) \propto R_{\rm sh}^{-1}$ (Eq.~\ref{eq:NH_R}) and, assuming a constant shock velocity, $N_H(t)\propto t^{-1}$.

The Thomson optical depth ahead of the shock is

\begin{equation}
\tau_T(r)
=
\int_r^{\infty}
\sigma_T n_e(r')\,dr'
=
\frac{\sigma_T}{\mu_e m_p}
\frac{\dot{M}}{4\pi v_w}
\frac{1}{r},
\end{equation}

where $\sigma_T$ is the Thomson cross section and
\begin{equation}
N_H(r)
=
\frac{\dot{M}}
{4\pi v_w \mu_e m_p r}
\end{equation}

according to Eq.\ref{eq:NH_R}, which implies  

\begin{equation}
\tau_T(r)
=
\sigma_T \mu_e N_H(r),
\end{equation}

Equivalently, the local CSM density can be written as

\begin{equation}
\rho_{\rm csm}(r)
=
\mu_e m_p
\frac{N_H(r)}
{r}.
\end{equation}

and, substituting this relation into the shock luminosity (Eq.  \ref{eq:L_sh}) with $R_{\rm sh}(t)=v_{\rm sh}t$, we find

\begin{equation}
L_{\rm sh}(t)
=
2\pi
\mu_e m_p
v_{\rm sh}^{4}
t
N_H(t).
\end{equation}

This result provides a direct connection between the observed column density and the shock luminosity under the assumptions of a steady wind and a constant shock velocity. 

We want to generalize this method allowing for an arbitrary CSM density profile and shock evolution in order to infer them directly from X-ray observations without assuming a specific wind profile.

\subsection{Data-driven reconstruction of the CSM density and shock dynamics}
\label{sec:data_driven}

We apply our data-driven framework to two nearby CSM-interacting supernovae, SN~2023ixf and SN~2024ggi, using their X-ray observations to reconstruct the CSM density profile and shock evolution.

SN~2023ixf, discovered in May 2023 in the nearby galaxy M101 \citep{Itagaki2023a,Itagaki2023b}, is one of the closest CCSN observed in recent years (redshift $z=0.000804$, distance $\sim6.7$ Mpc). Prediscovery observations indicate that the explosion occurred around 2023 May 18 \citep{Mao2023}.~\textit{Swift} began X-ray observations approximately 0.67 days after discovery (MJD = 60083.73), corresponding to about 1.6 days after first light \citep{Panjkov2023}. No soft X-ray emission in the 0.3-10 keV band was detected until about four days after first light, whereas hard X-ray emission above $\sim3$ keV was detected by \textit{NuSTAR} at $\sim3.9$ days \citep{Grefenstette2023}.

SN~2024ggi was discovered by the Asteroid Terrestrial-impact Last Alert System (ATLAS) on 2024 April 11 (MJD 60411.14) and subsequently classified as a hydrogen-rich CCSN \citep{Tonry2024,Srivastav2024,Chen_2025}.~Located in the nearby spiral galaxy NGC~3621 at a distance of $7.2\pm0.2$ Mpc and redshift $z=0.002215$ \citep{Saha_2006,JacobsonGalan2024a}, it provides an excellent second test case for our framework.

Our analysis is based on two observables derived from X-ray spectral fitting: the intrinsic hydrogen-equivalent column density, $N_{\rm H}$, which traces the amount of CSM located ahead of the shock, and the plasma temperature, $kT$, from which the shock velocity is inferred through the Rankine-Hugoniot relation.~Together, these observables allow us to reconstruct the shock evolution, infer the CSM density profile, estimate the SBO epoch, and predict the resulting high-energy neutrino flux.~For SN~2023ixf, we adopt the spectral parameters reported by \citet{Nayana_2025}, obtained from joint analyses of \textit{Swift}-XRT, \textit{NuSTAR}, \textit{Chandra}, and \textit{XMM-Newton}. For SN~2024ggi, we use the X-ray spectral measurements presented by \citet{Ferdinand_2026}, based on observations with \textit{Swift}-XRT, \textit{Chandra}, and \textit{XMM-Newton}. While uncertainties are available for both $N_{\rm H}$ and $kT$ in SN~2023ixf, the published temperature measurements of SN~2024ggi do not provide uncertainties at all epochs. Consequently, a complete uncertainty propagation can only be performed for SN~2023ixf. 

From the definition of the hydrogen-equivalent column density $N_{\rm H}$
(Eq.~\ref{eq:NH_R}) and using $v_{\rm sh}(t)=dR_{\rm sh}/dt$, 
the hydrogen number density at the shock position can be linked directly to the temporal evolution of the absorbing column as
\begin{equation}
n_H[R_{\rm sh}(t)]
=-\frac{1}{v_{\rm sh}(t)}
\frac{dN_H}{dt}.
\label{eq:nH1}
\end{equation}
Eq.~(\ref{eq:nH1}) is independent of any specific assumption about
the radial structure of the CSM or the temporal evolution of the shock
velocity.~In turn, the observed evolution of the hydrogen-equivalent column density is well described by a power-law (PL),
\begin{equation}
N_H(t)
=A\left(\frac{t}{t_{\rm ref}}\right)^{-\alpha},
\label{eq:NH}
\end{equation}
%where $t_{\rm ref}=1\,{\rm day}$ and $A=N_H(t_{\rm ref})$. 
where we adopt $t_{\rm ref}=1\,{\rm day}$ as a convenient reference timescale, so that the normalization $A$ corresponds directly to the
column density after one day, $A=N_H(t_{\rm ref})$. 
The parameters $A$ and $\alpha$ are determined directly from the X-ray measurements.

\begin{figure*}[ht]
\centering

\begin{minipage}[t]{0.48\textwidth}
    \centering
    \includegraphics[width=\textwidth]{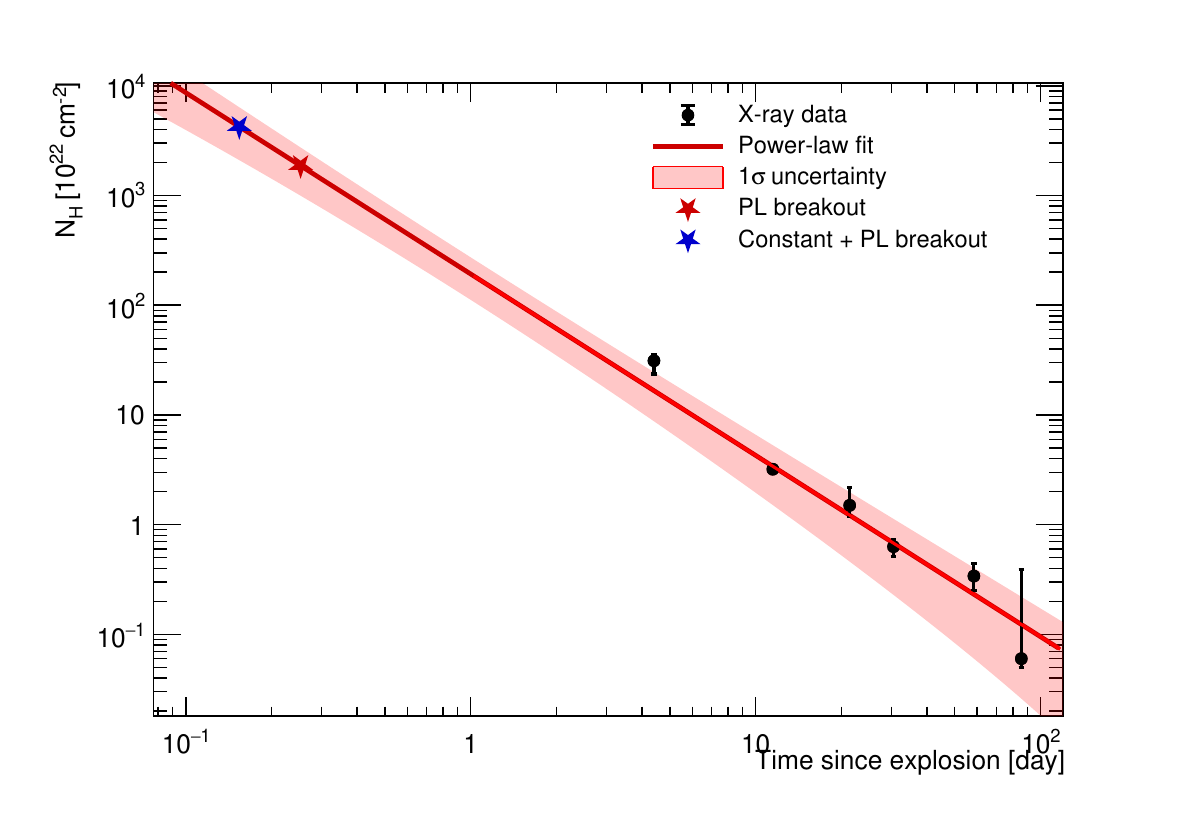}
    \textbf{(a)} SN~2023ixf
\end{minipage}
\hfill
\begin{minipage}[t]{0.48\textwidth}
    \centering
    \includegraphics[width=\textwidth]{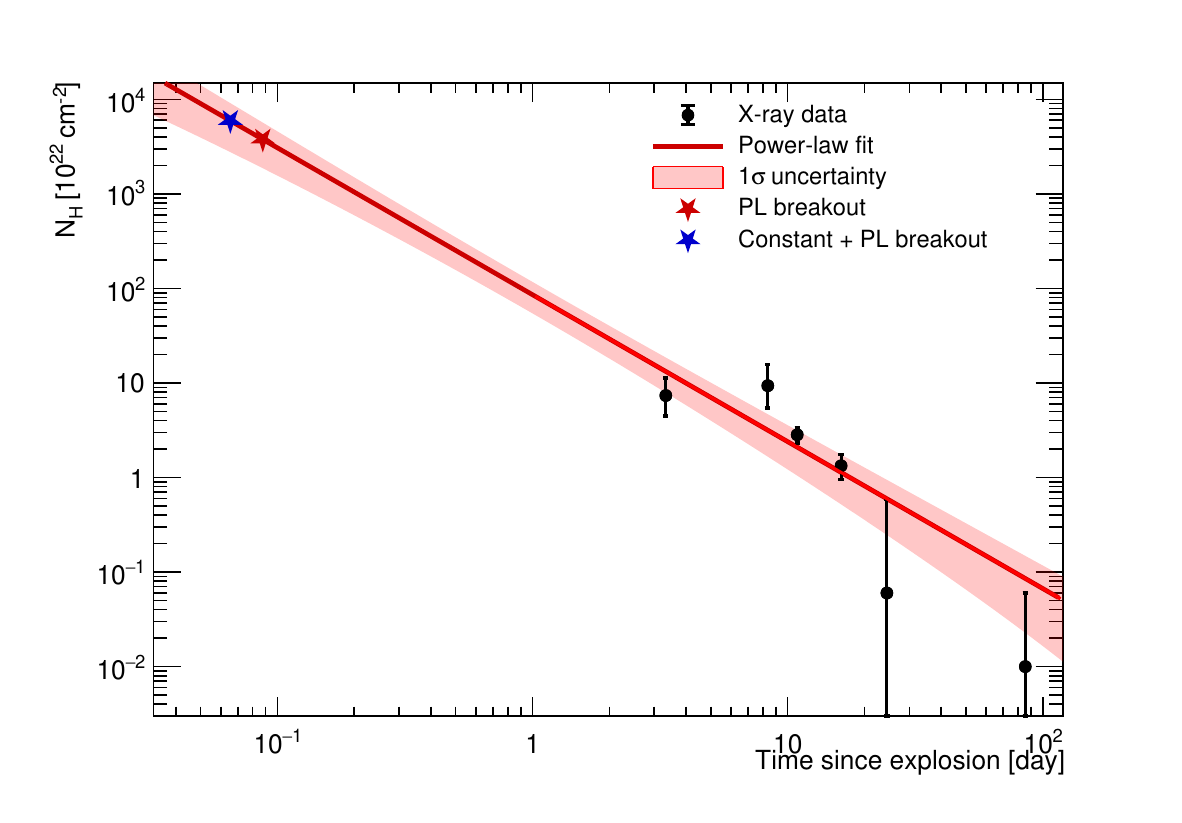}
    \textbf{(b)} SN~2024ggi
\end{minipage}

\caption{
Evolution of the hydrogen-equivalent column density $N_H$ for SN~2023ixf (a) and SN~2024ggi (b).
The solid lines show the best-fitting PL evolution adopted in the data-driven reconstruction.
}
\label{fig:NH_evolution}

\end{figure*}

Fig.~\ref{fig:NH_evolution} shows the temporal evolution of the intrinsic hydrogen-equivalent column density, $N_{\rm H}$, together with the best-fitting (PL) model. The second observable required by the reconstruction is the shock velocity, $v_{\rm sh}$, which is inferred from the measured plasma temperature using the Rankine-Hugoniot relation for strong shocks
\citep{Draine2011},
%\begin{equation} kT(t) =\frac{3}{16}\mu m_p v_{\rm sh}^2(t), \end{equation} which gives
\begin{equation}
v_{\rm sh}(t)
=\left[
\frac{16kT(t)}
{3\mu m_p}
\right]^{1/2}.
\label{eq:vsh_temperature}
\end{equation}

where $\mu$ is the mean molecular weight of the shocked plasma and $m_p$ is the proton mass. Empirically, the inferred shock velocities are well described as a PL decay,
\begin{equation}
v_{\rm sh}^{\rm PL}(t)
=
B\left(\frac{t}{t_{\rm ref}}\right)^{-\beta},
\label{eq:v_sh_PL}
\end{equation}
where $B=v_{\rm sh}(t_{\rm ref})$ and $\beta$ are determined by data fitting. This observational result, however, does not determine when the deceleration began. 

At sufficiently early times, before the swept-up CSM mass becomes dynamically important, the fastest outer ejecta are expected to undergo approximately free expansion, whereas interaction with a sufficiently dense CSM can lead to significant shock deceleration already at early times. The highest CSM densities and hadronic interaction efficiencies are reached at epochs earlier than the first X-ray velocity measurement. Thus, predicting the neutrino emission requires extrapolating the shock dynamics back to the observationally-unconstrained time interval between $t_{\rm bo}$ and $t_X$, the time first X-ray observation.

We therefore consider two limiting parameterizations for this early-time evolution. 
In the first case, labeled PL parameterization, we assume that the PL decay of $v_{\rm sh}$ was already established at the SBO epoch, and extrapolate the observed PL evolution back to $t_{\rm bo}$.
This corresponds to a scenario in which significant CSM-induced deceleration has already started during the earliest interaction phase.

In the second case, hereafter referred to as the
\textit{constant+power-law parameterization}(const+PL),
we adopt a more conservative scenario in which the shock undergoes approximately free expansion between $t_{\rm bo}$ and $t_X$, with a constant velocity equal to that measured at $t_X$, and subsequently follows the observed PL decay (Eq.~\ref{eq:v_sh_PL}), hence 

\begin{equation}
v_{\rm sh}^{\rm const+PL}(t)
=
\begin{cases}
v_{\rm sh}^{\rm PL}(t_X),
& t_{\rm bo}\leq t<t_X, \\[6pt]
B\left(\dfrac{t}{t_{\rm ref}}\right)^{-\beta},
& t\geq t_X.
\end{cases}
\label{eq:v_sh_constPL}
\end{equation}

\begin{figure*}[ht]
\centering

\begin{minipage}[t]{0.48\textwidth}
    \centering
    \includegraphics[width=\textwidth]{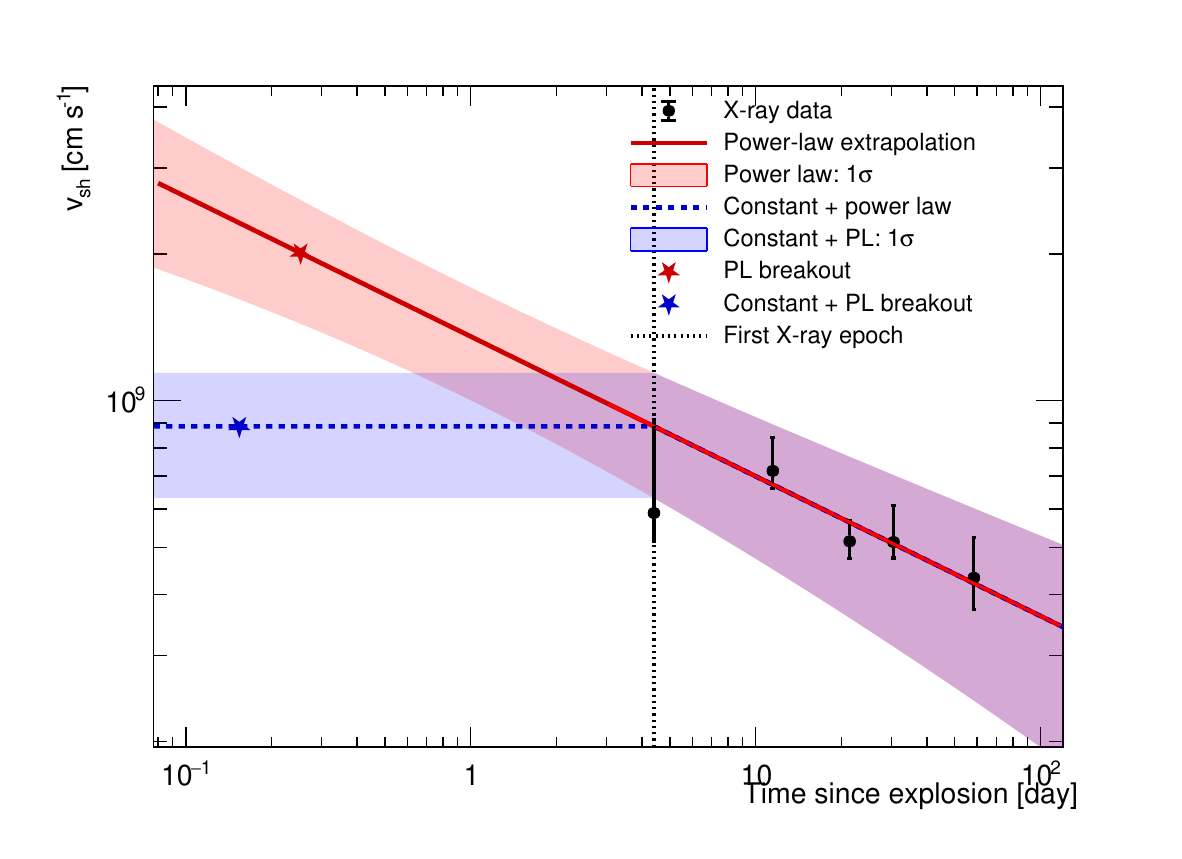}
    \textbf{(a)} SN~2023ixf
\end{minipage}
\hfill
\begin{minipage}[t]{0.48\textwidth}
    \centering
    \includegraphics[width=\textwidth]{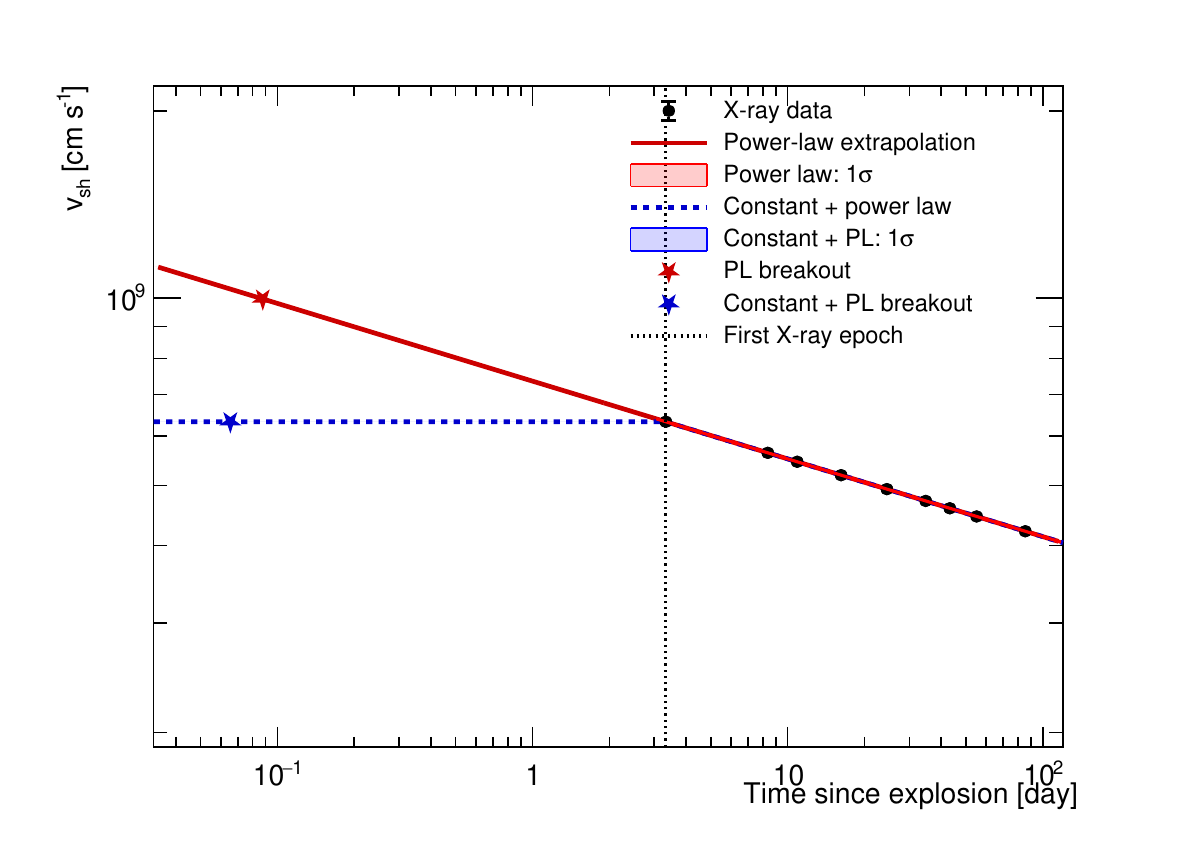}
    \textbf{(b)} SN~2024ggi
\end{minipage}

\caption{
Shock-velocity evolution adopted for SN~2023ixf (a) and SN~2024ggi (b). The points show the shock velocities inferred from the X-ray plasma temperature measurements. The solid curves show the best-fitting PL evolution extrapolated to epochs earlier than the first X-ray observation, while the dashed curves show the alternative const+PL parameterization. In the latter case, the
shock velocity is kept constant at $v_{\rm sh}(t_X)$ between shock breakout and the first X-ray epoch $t_X$, and follows the same fitted power law thereafter. The two parameterizations therefore differ only during the early-time interval not directly constrained by the X-ray velocity measurements.
}
\label{fig:vsh_evolution}

\end{figure*}

The two parameterizations are intended to bracket the uncertainty associated with the unobserved early-time shock evolution, rather than to provide a detailed dynamical solution.
Fig.~\ref{fig:vsh_evolution} shows the shock velocity, $v_{\rm sh}$, inferred from the X-ray plasma temperature of SN~2023ixf through the Rankine-Hugoniot relation. To test for evidence of shock deceleration, we fit the X-ray-derived velocities with both a constant-velocity model and the PL evolution given by Eq.~\ref{eq:v_sh_PL}. The constant-velocity fit yields $\chi^2/{\rm dof}=10.44/4$, while the PL fit gives $\chi^2/{\rm dof}=2.35/3$, with $\beta=0.287\pm0.083$. 

Since the constant-velocity model is recovered from the power-law model for $\beta=0$, the two models are nested. An $F$-test gives $F=10.33$ and $p=0.049$, providing evidence for a time-dependent shock velocity over the epochs sampled by the X-ray observations.

Importantly, this result constrains the shock dynamics only from the first X-ray epoch, $t_X$, onward. It therefore does not determine the shock evolution between breakout and $t_X$, nor when the inferred deceleration begins. We consequently consider the two early-time velocity parameterizations introduced in Section~\ref{sec:data_driven} for the interval $t_{\rm bo}\leq t<t_X$.

Figure~\ref{fig:vsh_evolution} shows the resulting PL and const+PL shock evolutions. Both parameterizations follow the same best-fitting power law over the X-ray-observed interval and differ only at earlier times.

In both velocity panels, the shaded regions represent the propagated $1\sigma$ uncertainties, while the star marks the shock-breakout epoch inferred from the condition $\tau_T \simeq c/v_{\rm sh}$ for the corresponding velocity parameterization.

The observed decline of both $N_{\rm H}$ and $v_{\rm sh}$ indicates that the supernova shock propagates through a CSM whose density decreases rapidly with radius. In particular, the evolution of the absorbing column density is significantly steeper than the $N_{\rm H}\propto t^{-1}$ behaviour expected for a steady stellar wind, providing the first observational indication that the CSM environment departs from the canonical wind scenario. The best-fit evolution of $N_{\rm H}(t)$ and $v_{\rm sh}(t)$ forms the observational basis for the reconstruction of the CSM density profile, the determination of the shock breakout epoch, and the prediction of the associated high-energy neutrino emission.

The shock radius corresponding to a given velocity parameterization is obtained kinematically as
\begin{equation}
R_{\rm sh}(t)
=R_{\rm bo}
+
\int_{t_{\rm bo}}^{t}
v_{\rm sh}(t'),dt',
\label{eq:Rsh_general}
\end{equation}
where $R_{\rm bo}=R_{\rm sh}(t_{\rm bo})$ is the shock radius at breakout. We assume that the shock propagation starts at the progenitor stellar radius, 
for which we adopt a fiducial stellar radius of $R_\star=3\times10^{13}$~cm ($\simeq430\,R_\odot$), within the broad range of $\sim100$--$1570\,R_\odot$ spanned by stellar evolution models during the red-supergiant phase \citep{Meynet2015}.
The breakout radius is obtained by integrating the adopted shock-velocity evolution from the stellar surface up to $t_{\rm bo}$.

The subsequent shock-radius evolution is calculated from Eq.~\ref{eq:Rsh_general} using either of the two velocity parameterizations introduced above. For the PL parameterization, the PL velocity evolution is integrated directly from the SBO epoch. For the const+PL parameterization, the shock expands at constant velocity $v_X$ between $t_{\rm bo}$ and $t_X$, and subsequently follows the power-law evolution inferred from the X-ray measurements. The explicit expressions for the shock radius in the two parameterizations are provided in Appendix~\ref{sec:6}. The resulting shock-radius evolution is used below to reconstruct the radial CSM density profile and to evaluate the shock dynamics and hadronic interaction efficiency.

The local hydrogen number density sampled by the shock follows directly from the temporal evolution of the absorbing column density through
Eq.~\ref{eq:nH1}. For the PL evolution of $N_H(t)$ given by Eq.~\ref{eq:NH}, and adopting the PL shock-velocity parameterization of
Eq.~\ref{eq:v_sh_PL}, we obtain
\begin{equation}
n_H^{\rm PL}(t)
=
\frac{\alpha A}{B t_{\rm ref}}
\left(\frac{t}{t_{\rm ref}}\right)^{-\alpha+\beta-1}.
\label{eq:nH_PL}
\end{equation}

For the const+PL parameterization, the reconstructed density is instead piecewise,
\begin{equation}
n_H^{\rm const+PL}(t)
=
\begin{cases}
\dfrac{\alpha A}{v_X t_{\rm ref}}
\left(\dfrac{t}{t_{\rm ref}}\right)^{-\alpha-1},
&
t_{\rm bo}^{\rm const+PL}\leq t<t_X,
\\[10pt]
\dfrac{\alpha A}{B t_{\rm ref}}
\left(\dfrac{t}{t_{\rm ref}}\right)^{-\alpha+\beta-1},
&
t\geq t_X,
\end{cases}
\label{eq:nH_constPL}
\end{equation}
where $v_X=v_{\rm sh}^{\rm PL}(t_X)$.

By construction, the two density reconstructions coincide from $t_X$ onward, since both velocity parameterizations follow the same observed PL evolution at these epochs. They differ only during the unobserved interval between shock breakout and the first X-ray velocity measurement. 
The PL extrapolation assumes that the decreasing velocity evolution inferred from the X-ray observations already applies at, or sufficiently close to, the SBO epoch. The const+PL parameterization instead provides a limiting case in which no additional velocity evolution is imposed between $t_{\rm bo}$ and $t_X$. These two parameterizations therefore bracket the uncertainty associated with the shock dynamics during the interval not directly constrained by the X-ray temperature measurements.

\subsection{Data-driven estimate of the shock breakout time}
\label{sec:breakout}

The SBO epoch is determined using the condition introduced in Eq.~\ref{eq:tau_T_t}, which marks the transition from the RMS to the CLS regime. Combining this condition with the observed evolution of the hydrogen-equivalent column density and the reconstructed shock velocity allows us to estimate $t_{\rm bo}$ directly from the X-ray-derived quantities.

Since the shock velocity is not directly constrained before the first X-ray measurement at $t_X$, the inferred breakout time depends on the adopted parameterization for the early-time shock evolution. For the PL parameterization, in which the X-ray-derived velocity evolution is extrapolated back to the breakout epoch, we obtain

\begin{equation}
t_{\rm bo}^{\rm PL}
=
t_{\rm ref}
\left(
\frac{\mu_e A B \sigma_T}{c}
\right)^{1/(\alpha+\beta)}.
\label{eq:tbo_PL}
\end{equation}

For the const+PL parameterization, the shock velocity between breakout and
the first X-ray velocity measurement is instead fixed at
$v_X=v_{\rm sh}^{\rm PL}(t_X)$. The corresponding breakout time is

\begin{equation}
t_{\rm bo}^{\rm const+PL}
=
t_{\rm ref}
\left(
\frac{\mu_e A v_X \sigma_T}{c}
\right)^{1/\alpha}.
\label{eq:tbo_constPL}
\end{equation}

The two estimates therefore quantify the uncertainty in the inferred SBO epoch associated with the unobserved early-time shock evolution. The corresponding breakout velocities are obtained directly from the respective velocity parameterizations evaluated at $t_{\rm bo}$.

\subsection{CSM density profile}
\label{sec:csm_density}

Using Eq.~\ref{eq:nH1}, the CSM mass density at the shock position can
be written in the general form
\begin{equation}
\rho(t)=
\mu_e m_p
\left[
-\frac{1}{v_{\rm sh}(t)}
\frac{dN_H}{dt}
\right].
\label{eq:rho_dndt}
\end{equation}

For the PL parametrization of the absorbing column density introduced above, Eq.~\ref{eq:NH}, the reconstructed CSM density depends on the adopted parameterization for the early-time shock velocity. Substituting Eq.~\ref{eq:v_sh_PL} into
Eq.~\ref{eq:rho_dndt} gives
\begin{equation}
\rho^{\rm PL}(t)
=
\mu_e m_p
\frac{\alpha A}{B\,t_{\rm ref}}
\left(
\frac{t}{t_{\rm ref}}
\right)^{-\alpha+\beta-1}.
\label{eq:rho_t_PL}
\end{equation}

For the const+PL parameterization, the reconstructed density is instead
piecewise,
\begin{equation}
\rho^{\rm const+PL}(t)
=
\begin{cases}
\displaystyle
\mu_e m_p
\frac{\alpha A}{v_X\,t_{\rm ref}}
\left(
\frac{t}{t_{\rm ref}}
\right)^{-\alpha-1},
&
t_{\rm bo}^{\rm const+PL}\leq t<t_X,
\\[12pt]
\displaystyle
\mu_e m_p
\frac{\alpha A}{B\,t_{\rm ref}}
\left(
\frac{t}{t_{\rm ref}}
\right)^{-\alpha+\beta-1},
&
t\geq t_X.
\end{cases}
\label{eq:rho_t_constPL}
\end{equation}

In both cases, the reconstructed CSM density profile depends only on the parameterized shock velocity evolution and on the \textit{observed} evolution of the X-ray absorbing column density, and temperature, without any assumption about pre-explosion mass-loss. Thus the two parameterizations for $v_{\rm sh}$ give the same CSM density from $t_X$ onward, while they differ at earlier times, where they are not directly constrained by the X-ray temperature measurements. % The corresponding radial density profile follows from the shock-radius evolution and therefore also depends on the adopted velocity parameterization.

For the PL parameterization, using the shock-radius evolution given by
Eq.~\ref{eq:Rsh_general}, the radial dependence of the reconstructed density can be written as
\begin{equation}
\rho^{\rm PL}(R_{\rm sh})
\propto
(R_{\rm sh}-R_0)^{-s_{\rm PL}},
\end{equation}
where
\begin{equation}
s_{\rm PL}
=
\frac{\alpha-\beta+1}{1-\beta}.
\label{eq:s_csm_PL}
\end{equation}
and $R_{\rm 0}$ is the initial radius at time $t_{\rm 0}$. Thus, for the PL parameterization, the radial slope of the
CSM density profile is determined entirely by the temporal slopes of
the observed absorbing column density and shock temperature, i.e. velocity (Eq.\ref{eq:vsh_temperature}).

For the const+PL parameterization, the shock radius evolves linearly during
the early-time interval,

\begin{equation}
R_{\rm sh}^{\rm const+PL}(t)
=
R_{\rm bo}^{\rm const+PL}
+
v_X
\left(t-t_{\rm bo}^{\rm const+PL}\right),
\qquad
t_{\rm bo}^{\rm const+PL}\leq t<t_X,
\label{eq:Rsh_constPL_density}
\end{equation}
while the density evolves as
\begin{equation}
\rho^{\rm const+PL}(t)
\propto t^{-\alpha-1}.
\label{eq:rho_constPL_scaling}
\end{equation}

The early-time radial density profile in the const+PL case is therefore
obtained by combining Eqs.~\ref{eq:Rsh_constPL_density} and
\ref{eq:rho_constPL_scaling}. In contrast to the PL case, the finite
values of $t_{\rm bo}^{\rm const+PL}$ and
$R_{\rm bo}^{\rm const+PL}$ introduce an offset between time and radius,
so that the early-time profile is not, in general, described by a
single power law in $(R_{\rm sh}-R_{\rm bo})$.

From $t_X$ onward, the const+PL parameterization follows the same velocity evolution as the PL parameterization. The two reconstructions therefore have the same local density evolution as a function of time and approach
the same radial slope,
\begin{equation}
s=
\frac{\alpha-\beta+1}{1-\beta}.
\end{equation}

\begin{figure}[t]
    \centering

    \begin{minipage}[t]{0.49\linewidth}
        \centering
        \includegraphics[width=\linewidth]{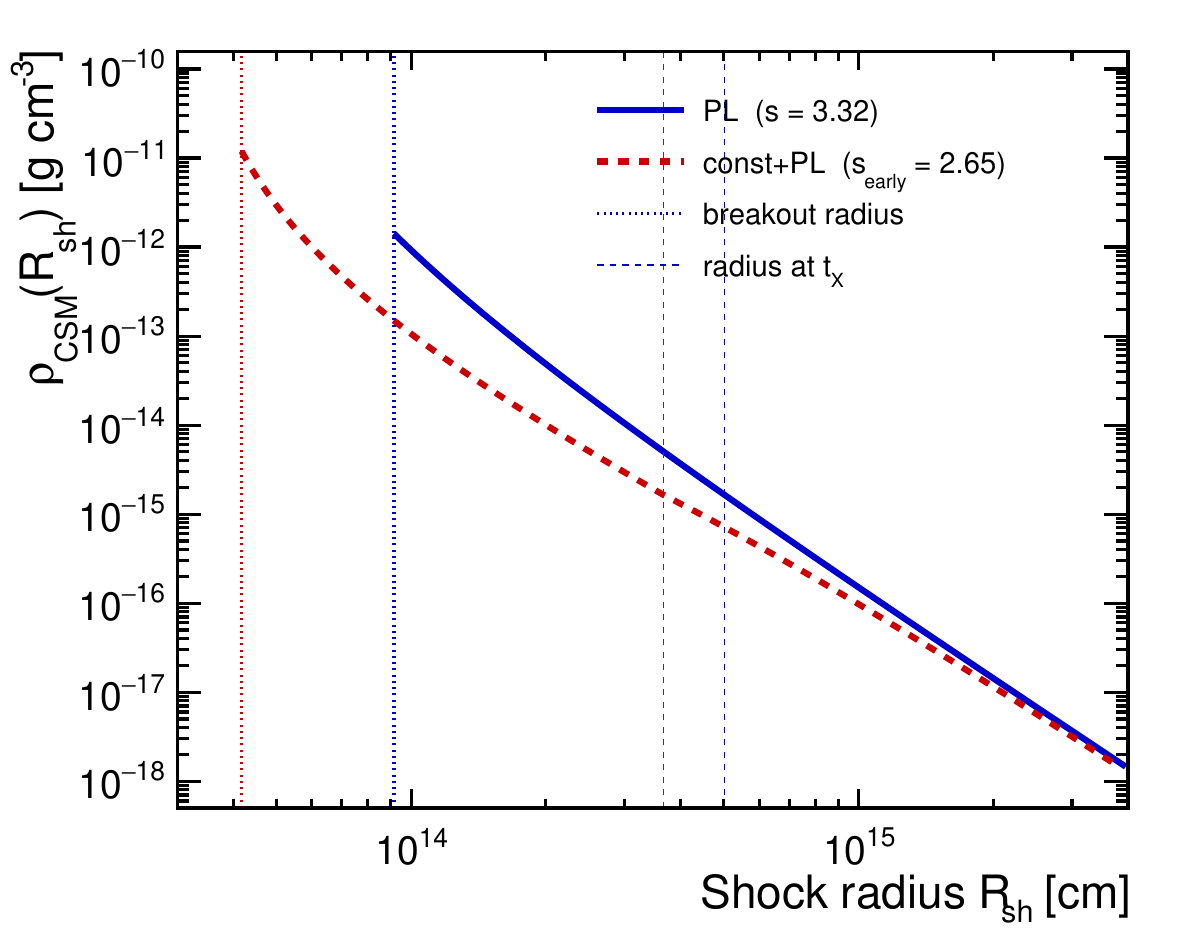}
        \textbf{(a)} SN~2023ixf
    \end{minipage}
    \hfill
    \begin{minipage}[t]{0.49\linewidth}
        \centering
        \includegraphics[width=\linewidth]{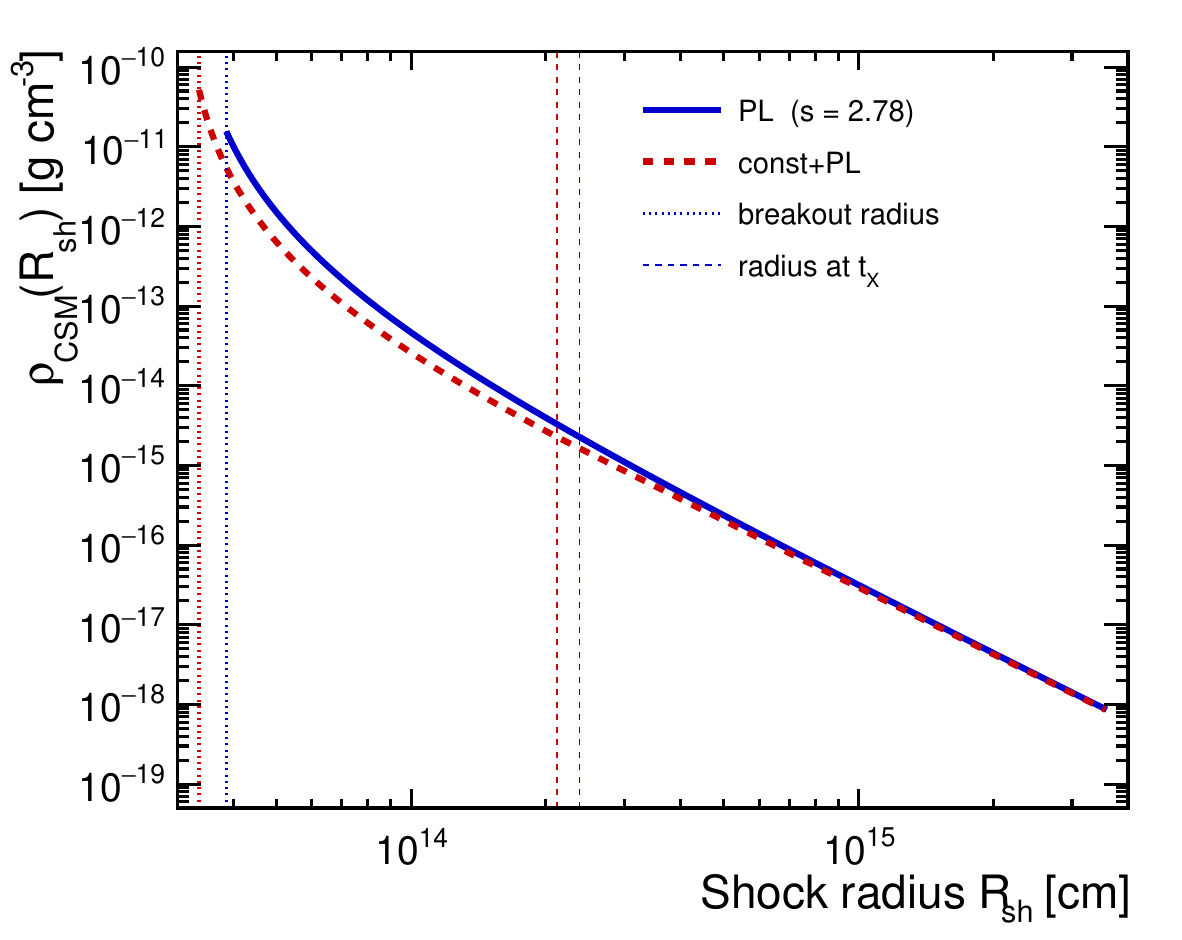}
        \textbf{(b)} SN~2024ggi
    \end{minipage}

    \caption{
    Reconstructed CSM density as a function of shock radius for
    SN~2023ixf and SN~2024ggi, for the PL and const+PL shock velocity
    parameterizations. The vertical lines indicate the shock radii
    corresponding to breakout and to the first X-ray temperature
    measurement, $t_X$, and are color-coded according to the two shock
    velocity parameterizations. For SN~2023ixf, the inferred breakout
    radii are $R_{\rm bo}=9.16\times10^{13}$~cm and
    $4.18\times10^{13}$~cm for the PL and const+PL prescriptions,
    respectively. For SN~2024ggi, the corresponding values are
    $R_{\rm bo}=3.86\times10^{13}$~cm and
    $3.36\times10^{13}$~cm.
    }
    \label{fig:rho_csm}
\end{figure}

The resulting radial density profiles are shown in
Fig.~\ref{fig:rho_csm}. This provides a direct connection between the X-ray observables and the underlying CSM structure, without requiring an a priori assumption about its radial profile. In particular, $s=2$ corresponds to the canonical steady-wind profile, whereas $s>2$ ($s<2$)
indicates a CSM density profile steeper (shallower) than a steady wind.

For SN~2023ixf, we infer a density profile approximately following $\rho_{\rm CSM}\propto (R_{\rm sh}-R_0)^{-3.3}$, while for SN~2024ggi we obtain a somewhat shallower profile,
$\rho_{\rm CSM}\propto (R_{\rm sh}-R_0)^{-2.89}$.
Thus, both reconstructed profiles are steeper than the canonical steady-wind profile, with SN~2023ixf showing the stronger deviation.
This suggests that the observed CSM cannot be adequately described by a stationary wind and instead points toward a time-dependent mass-loss history prior to core collapse.

\subsection{Hadronic interaction efficiency and neutrino luminosity}

High-energy neutrinos are produced through inelastic $pp$ interactions between accelerated CRs and the dense CSM. The $pp$ interaction timescale is
\begin{equation}
t_{pp}(t)=
\frac{1}
{\kappa_{pp}\,\sigma_{pp}\,n_{\rm CSM}(t)\,c},
\end{equation}
where $\sigma_{pp}\simeq5\times10^{-26}\ {\rm cm^2}$ is the inelastic $pp$ cross section, $\kappa_{pp}\simeq0.5$ is the inelasticity, and $n_{\rm CSM}(t)$ is the circumstellar hydrogen number density. The characteristic dynamical timescale of the shock is $t_{\rm dyn}(t)=R_{\rm sh}(t)/v_{\rm sh}(t)$, thus the efficiency of hadronic interactions is expressed as
\begin{equation}
f_{pp}(t)=
1-
\exp\left[
-\frac{t_{\rm dyn}(t)}
{t_{pp}(t)}
\right].
\end{equation}

Using Eq.~\ref{eq:nH1} the hadronic interaction efficiency eventually becomes
\begin{equation}
f_{pp}(t)=
1-
\exp\left[
-
\kappa_{pp}
\sigma_{pp}
c
\,
\frac{
R_{\rm sh}(t)
}
{
v_{\rm sh}^2(t)
}
\left(
-\frac{dN_H}{dt}
\right)
\right].
\label{eq:fpp_general}
\end{equation}

\begin{figure*}[ht]
\centering

\begin{minipage}[t]{0.49\textwidth}
\centering
\includegraphics[width=\linewidth]{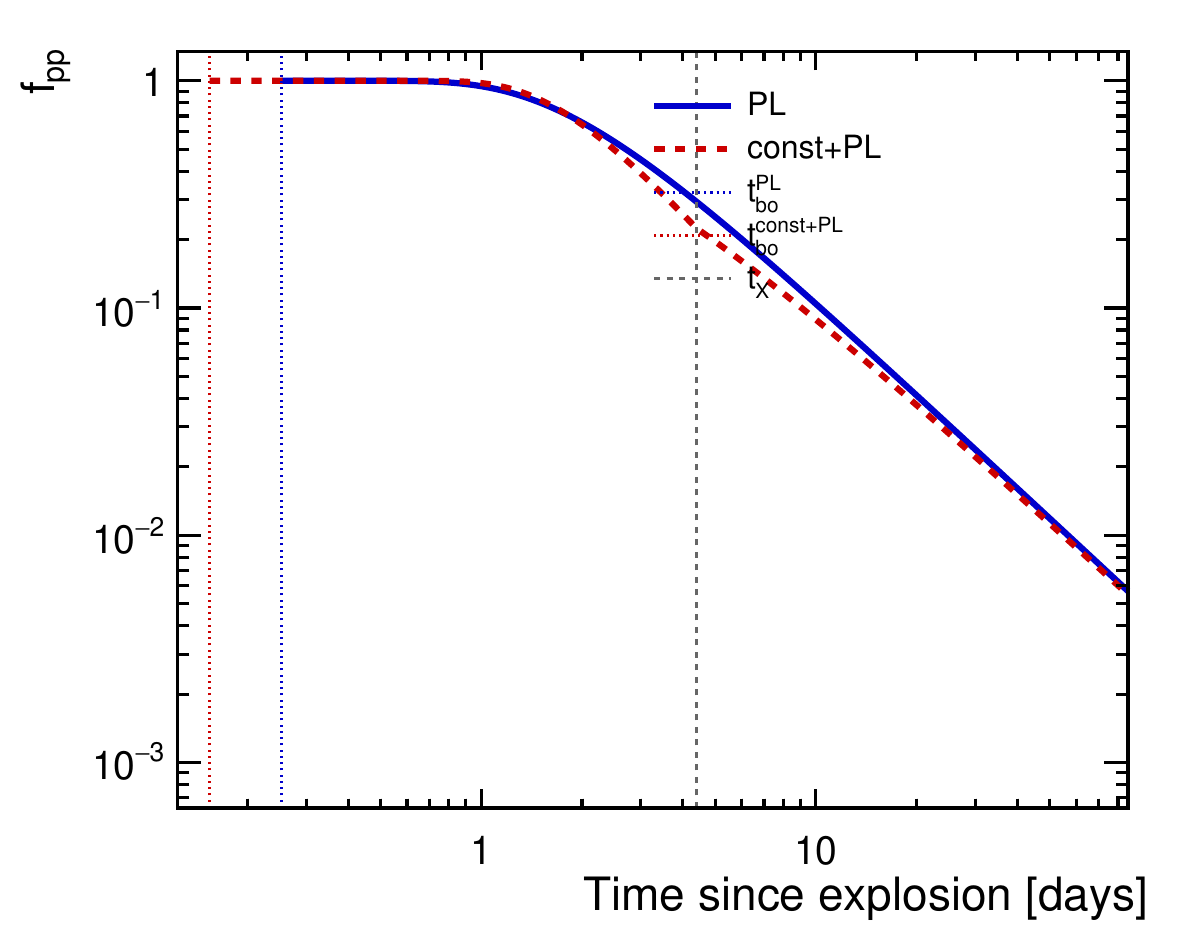}
\textbf{(a)} SN~2023ixf
\end{minipage}
\hfill
\begin{minipage}[t]{0.49\textwidth}
\centering
\includegraphics[width=\linewidth]{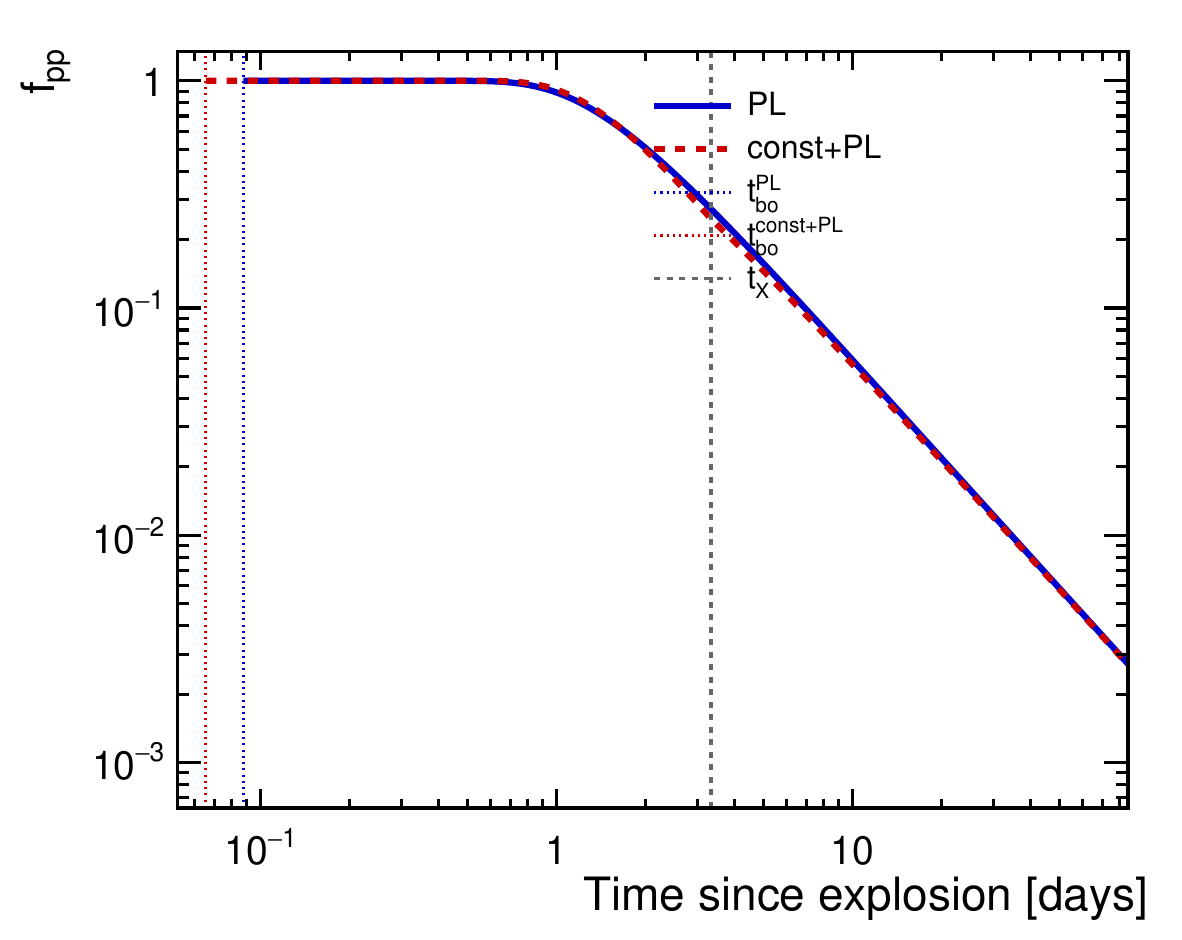}
\textbf{(b)} SN~2024ggi
\end{minipage}

\caption{
Time evolution of the $pp$ interaction efficiency, $f_{pp}$, for
SN~2023ixf (left) and SN~2024ggi (right), computed using the fitted
evolution of the X-ray absorbing column density and shock velocity.
The vertical dashed line marks the data-driven shock breakout epoch,
$t_{\rm bo}$, estimated from the condition
$\tau_T \simeq c/v_{\rm sh}$. In both sources, $f_{pp}$ is largest
at early times, when the CSM column density is highest, and progressively
decreases as the shock expands through the surrounding medium.
The higher values obtained for SN~2023ixf reflect its denser CSM
environment.
}
\label{fig:fpp}

\end{figure*}

Using the reconstructed CSM density and shock evolution, we compute the pp interaction efficiency. The results are shown in Figure~\ref{fig:fpp}. In both supernovae, $f_{pp}$ reaches its maximum value close to shock breakout, when the circumstellar column density is highest. SN~2023ixf exhibits systematically larger values of $f_{pp}$ than SN~2024ggi, indicating a denser environment and a more efficient conversion of CR energy into secondary particles and neutrinos.

Substituting the data-driven reconstructed density (Eq.~\ref{eq:rho_dndt}) into Eq.~\ref{eq:L_sh} for the shock power, we obtain
\begin{equation}
L_{\rm sh}(t)=
2\pi
\mu_e m_p
R_{\rm sh}^2(t)
v_{\rm sh}^2(t)
\left(
-\frac{dN_H}{dt}
\right) \, 
\label{eq:Lsh_general}
\end{equation}

and, recalling that the neutrino luminosity is $L_\nu(t)=\epsilon_{\rm CR}\, f_{pp}(t)\, L_{\rm sh}(t)$, we obtain the general time-dependent neutrino luminosity
\begin{equation}
L_\nu(t)=
\epsilon_{\rm CR}
\left[
1-
\exp\left(
-
\kappa_{pp}
\sigma_{pp}
c
\,
\frac{
R_{\rm sh}(t)
}
{
v_{\rm sh}^2(t)
}
\left(
-\frac{dN_H}{dt}
\right)
\right)
\right]
\,
2\pi
\mu_e m_p
R_{\rm sh}^2(t)
v_{\rm sh}^2(t)
\left(
-\frac{dN_H}{dt}
\right)\, .
\label{eq:Lnu_general}
\end{equation}

\begin{figure*}[ht]
\centering

\begin{minipage}[t]{0.49\textwidth}
\centering
\includegraphics[width=\linewidth]{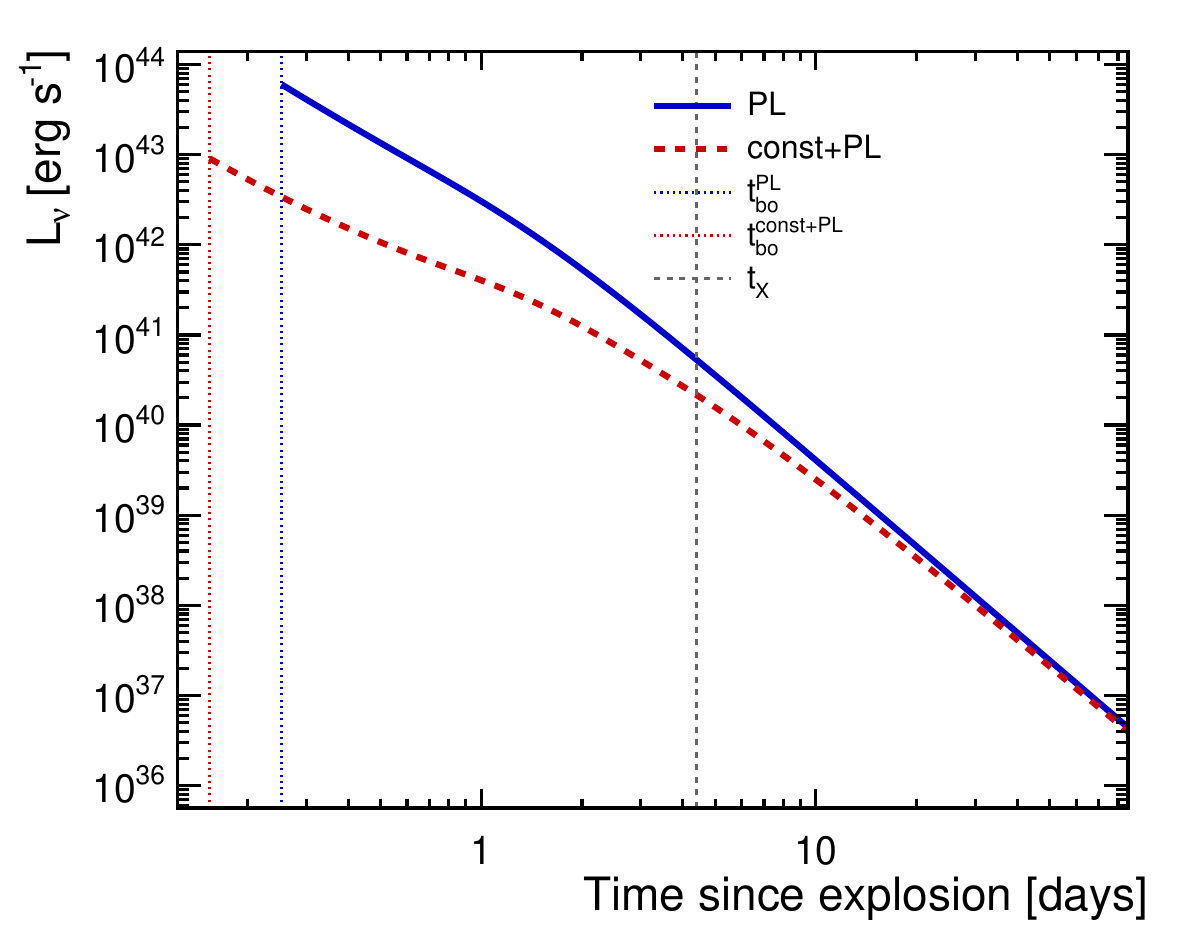}
\textbf{(a)} SN~2023ixf
\end{minipage}
\hfill
\begin{minipage}[t]{0.49\textwidth}
\centering
\includegraphics[width=\linewidth]{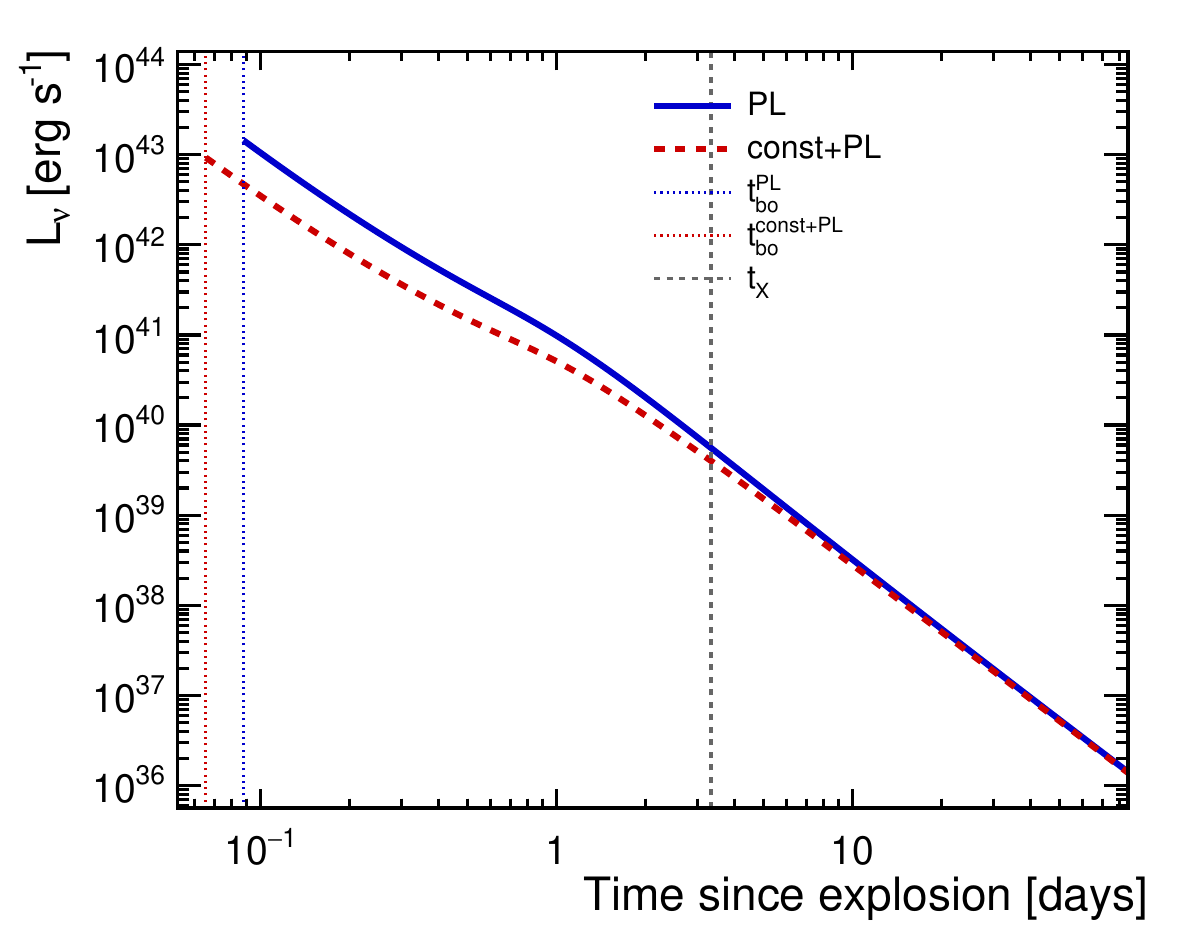}
\textbf{(b)} SN~2024ggi
\end{minipage}

\caption{
Time evolution of the neutrino luminosity, $L_{\nu}(t)$, for
SN~2023ixf (left) and SN~2024ggi (right). The luminosity is derived
from the observed evolution of the X-ray absorbing column density and
shock velocity, assuming hadronic neutrino production through $pp$
interactions. In both cases, the neutrino emission peaks shortly after
shock breakout and subsequently declines as the circumstellar density
decreases. The larger luminosity predicted for SN~2023ixf reflects its
denser circumstellar environment and higher hadronic interaction
efficiency.
}
\label{fig:Lnu}

\end{figure*}

These expressions are general and can be evaluated using either of the two parameterizations adopted for the early-time shock evolution. Equation~\ref{eq:Lnu_general} therefore relates the neutrino luminosity directly to the observed evolution of the X-ray absorbing column density and to the reconstructed shock radius and velocity. It can be evaluated for any parameterization adopted for the early-time shock evolution. The explicit expressions obtained for the different shock-velocity parameterizations considered in this work are derived in Appendix~\ref{sec:6}. The resulting time evolution of the neutrino luminosity for SN~2023ixf and SN~2024ggi is shown in
Fig.~\ref{fig:Lnu}.

The resulting neutrino luminosities are shown in Figure~\ref{fig:Lnu}. In both cases, the emission is strongly concentrated at early times, reflecting the combined effect of a high shock power and efficient hadronic interactions. As the CSM density decreases, the interaction efficiency drops and the neutrino luminosity rapidly declines. Owing to its denser CSM environment, SN~2023ixf is predicted to produce a substantially larger neutrino output than SN~2024ggi.

\section{Neutrino spectrum and detectability}
\label{sec:4}

The neutrino luminosity is related to the differential neutrino
production rate through
\begin{equation}
L_\nu(t)
=
\int_{E_{\min}}^{E_{\max}}
E_\nu
\frac{d\dot{N}_\nu}{dE_\nu}(E_\nu,t)
\,dE_\nu .
\label{eq:Lnu-def}
\end{equation}

We assume an $E_\nu^{-2}$ neutrino spectrum between $E_{\min}$ and
$E_{\max}$. Normalizing the spectrum to the neutrino luminosity defined
in Eq.~\ref{eq:Lnu-def} gives
\begin{equation}
E_\nu^2\frac{d\dot{N}_\nu}{dE_\nu}(E_\nu,t)
=
\frac{L_\nu(t)}
{\ln(E_{\max}/E_{\min})}.
\label{eq:nu_spectrum}
\end{equation}

The time-integrated neutrino spectrum is therefore
\begin{equation}
E_\nu^2\frac{dN_\nu}{dE_\nu}
=
\frac{1}{\ln(E_{\max}/E_{\min})}
\int_{t_0}^{t}
L_\nu(t')\,dt' .
\label{eq:nu_spectrum_integrated}
\end{equation}

The expected number of detected neutrino events is obtained by folding the time-integrated neutrino fluence at Earth with the detector effective area,
\begin{equation}
N_\nu =
\int
\frac{dN_\nu}{dE_\nu dS}
A_{\rm eff}(E_\nu,\delta)\,dE_\nu ,
\label{eq:Nevents}
\end{equation}
where $A_{\rm eff}(E_\nu,\delta)$ accounts for the energy and directional response of the detector. In this work, we consider the effective areas of IceCube and KM3NeT/ARCA \citep{art:icecube_aeff,art:loi}.
The detectability of the signal is determined by the atmospheric neutrino background expected within the same temporal and angular search window. The atmospheric neutrino background spectrum is taken from \citet{art:atmosferici}, and the calculation of the expected signal and background event rates follows the procedure described in \citep{Gagliardini2024}. For IceCube, the declination dependence is included directly through the detector effective area, whereas for KM3NeT/ARCA the source visibility is accounted for through the visibility function $V(\delta)$ \citep{Aiello2019}. Throughout this work, we adopt a conservative search cone with a half-opening angle of $5^\circ$. The expected background therefore scales linearly with the duration $T$ of the adopted search window.

\subsection{Optimal neutrino search window}
\label{sec:optimal_window}

The neutrino emission predicted by our model is strongly concentrated at early times after SBO, when the shock power is high and hadronic interactions are most efficient. Although the total number of expected neutrino events increases when integrating over longer time intervals, the atmospheric background also increases with time. Therefore, extending the search window does not necessarily improve the detectability of the neutrino signal.

The total number of signal events accumulated within a search window of duration $T$, starting at SBO, is then
\begin{equation}
N_\nu(T) =
\int_{t_{\rm bo}}^{t_{\rm bo}+T}
\dot{N}_\nu(t)\,dt .
\end{equation}
where $\dot{N}_\nu(t)$ represents the number of neutrino events expected per unit time.

Assuming an approximately constant atmospheric background rate, the
number of background events accumulated over the same interval increases
linearly with the duration of the search window, $N_{\rm bg}(T) \propto T$. We therefore define a simple quantity that describes the balance between the accumulated neutrino signal and the corresponding atmospheric background,
\begin{equation}
\mathcal{Q}(T)
=
\frac{N_\nu(T)}{\sqrt{N_{\rm bg}(T)}}
\propto
\frac{N_\nu(T)}{\sqrt{T}}.
\end{equation}

At early times, the neutrino signal accumulates rapidly, while the
contribution from the atmospheric background remains limited.
As the neutrino production efficiency decreases (Fig.~\ref{fig:fpp}),
the signal accumulation slows down, whereas the background continues
to grow with the duration of the search window.

We define $t_{\rm best}$ as the duration of the search window that
maximizes $\mathcal{Q}(T)$, i.e.
\begin{equation}
t_{\rm best}
=
\operatorname*{arg\,max}_{T}
\mathcal{Q}(T).
\end{equation}

For each shock-evolution parameterization and detector, we calculate
$\mathcal{Q}(T)$ as a function of the integration time and determine the
corresponding value of $t_{\rm best}$. The resulting optimal search windows,
together with the expected signal and atmospheric background events for
SN~2023ixf and SN~2024ggi, are reported in Table~\ref{tab:detectability}.
In all cases, the signal and background event numbers are integrated over
the interval $[t_{\rm bo},\,t_{\rm bo}+t_{\rm best}]$.

\begin{table}[b]
\centering
\caption{
Optimal neutrino search windows and corresponding expected signal and
atmospheric background events for SN~2023ixf and SN~2024ggi under the two
parameterizations adopted for the early-time shock evolution.
The quantity $t_{\rm best}$ denotes the duration of the search window,
starting at shock breakout, that maximizes
$\mathcal{Q}=N_\nu/\sqrt{N_{\rm bg}}$.
Both signal and background are integrated over
$[t_{\rm bo},\,t_{\rm bo}+t_{\rm best}]$.
The false-alarm probability is evaluated for the observation of at least
one background event,
$p_{\rm FA}=1-\exp(-N_{\rm bg})$.
The last column gives the probability that an observed event is associated
with the signal, conditional on exactly one event being detected,
$P(S|1)=N_\nu/(N_\nu+N_{\rm bg})$.
}
\label{tab:detectability}

\begin{tabular}{llcccccc}
\hline
Model & Detector &
$t_{\rm best}$ [d] &
$N_\nu$ &
$N_{\rm bg}$ &
$\mathcal{Q}_{\max}$ &
$p_{\rm FA}$ &
$P(S|1)$ \\
\hline

\multicolumn{8}{c}{SN~2023ixf} \\
\hline
PL & IceCube & 0.223 & 0.262 & 0.0218 & 1.78 & 0.0215 & 0.923 \\
PL & KM3NeT/ARCA & 0.221 & 0.121 & 0.00798 & 1.36 & 0.00795 & 0.938 \\
\hline
const+PL & IceCube & 0.172 & 0.0289 & 0.0167 & 0.224 & 0.0166 & 0.634 \\
const+PL & KM3NeT/ARCA & 0.172 & 0.0135 & 0.00620 & 0.171 & 0.00618 & 0.685 \\
\hline

\multicolumn{8}{c}{SN~2024ggi} \\
\hline
PL & IceCube & 0.0705 & $1.82\times10^{-3}$ & $2.80\times10^{-6}$ & 1.08 & $2.80\times10^{-6}$ & 0.998 \\
PL & KM3NeT/ARCA & 0.0705 & 0.0218 & 0.00686 & 0.263 & 0.00684 & 0.761 \\
\hline
const+PL & IceCube & 0.0547 & $9.01\times10^{-4}$ & $2.17\times10^{-6}$ & 0.611 & $2.17\times10^{-6}$ & 0.998 \\
const+PL & KM3NeT/ARCA & 0.0547 & 0.0108 & 0.00533 & 0.148 & 0.00531 & 0.670 \\
\hline

\hline
\end{tabular}
\end{table}

Once the optimal search window has been identified, we evaluate the statistical significance of a possible neutrino detection within this interval. Since the expected number of atmospheric background events can be much smaller than unity, the Gaussian approximation is not appropriate and we instead use Poisson statistics.

For an observed number of events $n_{\rm obs}$, the false-alarm probability (FAP) is defined as the probability that the atmospheric background alone produces $n_{\rm obs}$ or more events within the selected temporal and angular window,
\begin{equation}
p_{\rm FA}
=
P(N\geq n_{\rm obs}\mid N_{\rm bg})
=
\sum_{k=n_{\rm obs}}^{\infty}
\frac{N_{\rm bg}^{\,k}e^{-N_{\rm bg}}}{k!}.
\label{eq:FAP}
\end{equation}

In particular, for the observation of at least one neutrino event,
$n_{\rm obs}=1$, this reduces to
\begin{equation}
p_{\rm FA}
=
1-e^{-N_{\rm bg}}.
\end{equation}

The false-alarm probability therefore quantifies the probability that
an observed neutrino event within the optimal search window is produced
by the atmospheric background alone. Smaller values of $p_{\rm FA}$
correspond to a less likely background fluctuation and hence to a more
significant neutrino association.

In addition to the false-alarm probability, we quantify the probability
that an observed event is associated with the astrophysical signal.
Assuming that the signal and atmospheric background are independent
Poisson processes with expected event numbers $N_\nu$ and $N_{\rm bg}$,
respectively, if exactly one event is observed within the search window,
the probability that this event originates from the signal is

\begin{equation}
P({\rm signal}\mid 1)
=
\frac{N_\nu}
{N_\nu+N_{\rm bg}}.
\label{eq:signal_probability}
\end{equation}

Within the assumptions adopted in our model, we can estimate the distance
at which a SN~2023ixf-like source would yield one expected neutrino event
within the optimal search window. Since the expected number of signal
events scales as $N_\nu \propto D^{-2}$, this distance is given by

\begin{equation}
D_{1\nu} = D_0 \sqrt{N_\nu(D_0)},
\end{equation}

where $D_0=6.7$~Mpc is the distance adopted for SN~2023ixf. For the PL parameterization, we obtain $D_{1\nu}\simeq3.43$~Mpc for IceCube and $2.33$~Mpc for KM3NeT/ARCA, while for the const+PL parameterization the corresponding distances are $1.14$~Mpc and $0.78$~Mpc, respectively.
We further define a characteristic detection distance by requiring simultaneously $N_\nu\geq1$ and $\mathcal{Q}=N_\nu/\sqrt{N_{\rm bg}}\geq3$ within the optimal search window. Since the atmospheric background within the same temporal and angular window is independent of the source distance, both $N_\nu$ and $\mathcal{Q}$ scale as $D^{-2}$. The condition $\mathcal{Q}=3$ corresponds to distances of approximately $5.16$ and $4.51$~Mpc for IceCube and KM3NeT/ARCA, respectively, for the PL parameterization, and $1.83$ and $1.60$~Mpc  for the const+PL parameterization. In all cases, these distances are larger than the corresponding $D_{1\nu}$ values. Therefore, the requirement $N_\nu\geq1$ sets the limiting distance, yielding $D_{\max}=D_{1\nu}$ for all the cases considered here.

\section{Discussion and conclusions}
\label{sec:5}

In this work, we have developed a data-driven framework based on X-ray observations of CCSNe interacting with a dense CSM, which allows us to reconstruct the properties of the CSM and assess their implications for high-energy neutrino production.
By combining the temporal evolution of the hydrogen-equivalent column density, $N_{\rm H}(t)$, with the shock
velocity inferred from the X-ray temperature, the method provides a direct connection between X-ray observables, the radial CSM structure, the shock evolution, and the expected neutrino emission.

\paragraph{Data-driven reconstruction of the CSM density profile.}

A main advantage of the proposed approach lies in the direct reconstruction of the CSM density profile from the observed evolution of $N_{\rm H}$ and $v_{\rm sh}$, without assuming a steady-wind profile a priori. As the shock propagates outward, the decrease in $N_{\rm H}$ traces the variation of the absorbing material ahead of the shock. Combined with the shock velocity, this evolution provides the local CSM density as a function of the shock radius.

For both SN~2023ixf and SN~2024ggi, the reconstructed CSM profiles are steeper than the canonical $\rho\propto r^{-2}$ profile expected for a steady stellar wind. We infer a radial slope of $s\simeq3.3$ for SN~2023ixf and $s\simeq2.89$ for SN~2024ggi. Thus, while both objects show a significant departure from a steady-wind profile, the CSM reconstructed for SN~2023ixf is somewhat steeper than that of SN~2024ggi. In both cases, the steep density profiles imply that the density of target material encountered by the shock decreases rapidly as the shock propagates outward.

This behavior has direct consequences for high-energy neutrino production. As the CSM density decreases, the $pp$ interaction timescale increases and the corresponding hadronic interaction efficiency, $f_{pp}$, declines. 

Consequently, the neutrino luminosity is highest during the early phases of the interaction, when the CSM density is higher and hadronic interactions are more efficient. The X-ray-derived density reconstruction therefore provides not only information on the structure of the CSM, but also a direct observational handle on the temporal evolution of the conditions relevant for high-energy neutrino production.

\paragraph{Early-time shock evolution and systematic uncertainties.}

An important uncertainty in the framework arises from the extrapolation of the X-ray-derived evolution to epochs earlier than the first available X-ray observation where no direct measurements of \(N_{\rm H}\) and \(v_{\rm sh}\) are available. Both the determination of the SBO epoch and the predicted early-time neutrino emission require extending the inferred $N_{\rm H}(t)$ and $v_{\rm sh}(t)$ relations toward earlier times.

While extrapolating $N_{\rm H}(t)$ assumes that the reconstructed CSM structure persists toward smaller radii, the early-time evolution of the shock velocity depends on the poorly constrained dynamics between SBO and the first X-ray measurement. We have therefore considered different parameterizations for the early-time shock evolution. The resulting differences provide an estimate of the systematic uncertainty associated with this unobserved phase.

For SN~2023ixf, the absolute normalization of the predicted neutrino signal is also mildly sensitive to the assumed stellar radius, which sets the inner radial scale of the interaction. Varying the stellar radius between $R_\star=10^{13}$ and $10^{14}$~cm changes the expected neutrino event yield by approximately a factor of $3.5$ in our calculation. For our
fiducial parameters, the predicted number of detected events is generally below unity, but remains non-negligible. In particular, configurations toward the upper end of the considered stellar radius range can increase the IceCube expectation to values of order unity in the most favorable cases. Thus, although a single detected neutrino is not the expectation for the fiducial model, the probability of obtaining a neutrino detection is not negligible.
The detection prospects also depend strongly on the source distance. For a SN~2023ixf-like event, our criteria ($N_\nu\geq1$ and $\mathcal{Q}\geq3$) correspond to limiting distances of approximately $3.4$~Mpc for IceCube and $2.3$~Mpc for KM3NeT/ARCA in the more favorable PL scenario. In both cases, the limiting condition is set by the requirement of at least one expected
signal event.\\
The detection prospects could be further improved by next-generation neutrino observatories such as IceCube-Gen2, whose increased effective area is expected to enhance the event yield from transient sources.

The predicted event yield also scales with the assumed fraction of shock power transferred to accelerated CRs. Throughout this work we adopt the fiducial value $\epsilon_{\rm CR}=0.1$. This is a conservative choice, since a larger acceleration efficiency would increase the neutrino luminosity and the corresponding expected number of detected events.  The predicted number of neutrino events therefore depends not only on the X-ray derived properties of the CSM, but also on the assumptions adopted for the early-time shock evolution and the CR acceleration efficiency.

Despite all these uncertainties, the different parametrization adopted for the unobserved early-time shock evolution yield relatively similar SBO epochs. This is particularly relevant for neutrino searches, since the phase immediately following SBO corresponds to the highest CSM densities and therefore to the most favorable conditions for efficient hadronic interactions. Thus, while the predicted number of events is generally below unity and remains subject to systematic uncertainties, the model consistently identifies the early post-breakout phase as the most promising temporal window for a neutrino search.

\paragraph{Optimal temporal window for neutrino searches.}

The rapid decline of the neutrino luminosity has an important consequence for time-dependent searches. Most of the expected neutrino signal is accumulated during the early post-breakout evolution, whereas the atmospheric neutrino background increases approximately linearly with the duration of the search window. As the neutrino luminosity declines, the cumulative signal increases progressively more slowly, while the background continues to accumulate.
We therefore define $t_{\rm best}$ as the integration time, starting from SBO, that provides the most favorable balance between accumulating neutrino signal and limiting the atmospheric background. Beyond this optimal interval, the additional signal becomes increasingly small compared with the corresponding increase in background. For the cases investigated here, $t_{\rm best}$ lies within the early post-breakout evolution and is only moderately affected by variations in the early-time shock parameterization and stellar radius.

\paragraph{Statistical interpretation of a neutrino candidate.}

The small number of expected atmospheric background events within these short temporal windows makes Poisson statistics more appropriate than a Gaussian approximation for evaluating a possible neutrino association.

For SN~2023ixf, under the PL parameterization, the expected atmospheric background within the optimal search window is $N_{\rm bg}=0.0218$ for IceCube and $N_{\rm bg}=0.00798$ for KM3NeT/ARCA. The corresponding FAP for the observation of at least one neutrino are $p_{\rm FA}=2.15\%$ and $0.80\%$, respectively. Conditional on
exactly one event being detected, the probability that it is associated with the signal is $P(S|1)=92.3\%$ for IceCube and $93.8\%$ for KM3NeT/ARCA.

For the const+PL prescription, the lower predicted signal yields $P(S|1)=63.4\%$ and $68.5\%$ for IceCube and KM3NeT/ARCA, respectively, while the corresponding FAP remain small, $p_{\rm FA}=1.66\%$ and $0.62\%$. These results illustrate that a low FAP alone does not necessarily imply a strong association with the source, since the
interpretation of a neutrino candidate also depends on the expected signal. The statistical interpretation of a single neutrino candidate therefore remains sensitive to the assumed early-time shock evolution.

\paragraph{Multi-messenger implications.}

Interestingly, the early post-breakout phase that is most favorable for neutrino production may also be the least directly accessible through X-ray observations. At these epochs, the large CSM column density can strongly attenuate the X-ray emission produced by the shock, while the absorbed radiation may be reprocessed and emerge at UV and optical wavelengths. Rapid UV observations, such as those expected with ULTRASAT\citep{Shvartzvald_2024}, may therefore provide an important probe of SBO and of the earliest interaction between the ejecta and the surrounding CSM.

High-energy neutrinos provide a complementary probe of the early post-breakout phase. Unlike electromagnetic radiation, neutrinos escape essentially unaffected from the dense CSM. A neutrino detected within the physically motivated early-time window could therefore probe efficient hadronic interactions at epochs when the shock is still difficult to observe directly in X-rays.

More generally, our results show that, without requiring a detailed model of the CSM, the X-ray evolution of interacting supernovae can provide direct information on the CSM properties and on the temporal evolution relevant for high-energy neutrino production. Combining early X-ray and UV observations with time-dependent neutrino searches therefore provides a multi-messenger approach to probe the early phases of SN-CSM interaction.

\section*{Acknowledgements}

The authors thank Debjit Chatterjee for helpful discussions and valuable insights into the interpretation of the X-ray observations.

\appendix

\section{Analytical expressions for the neutrino luminosity}
\label{sec:6}

In this Appendix, we provide the explicit expressions for the neutrino
luminosity corresponding to the two parameterizations adopted for the
early time shock velocity evolution. In both cases, the shock radius is
obtained from the general relation given by Eq.~\ref{eq:Rsh_general}.

\subsection{Power-law extrapolation}

In the PL scenario, the shock velocity inferred from the X-ray
observations is extrapolated back to the SBO epoch according to
Eq.~\ref{eq:v_sh_PL}. Substituting the PL velocity evolution into
Eq.~\ref{eq:Rsh_general} gives

\begin{equation}
R_{\rm sh}^{\rm PL}(t)
=
R_{\rm bo}^{\rm PL}
+
\frac{B t_{\rm ref}}{1-\beta}
\left[
\left(\frac{t}{t_{\rm ref}}\right)^{1-\beta}
-
\left(\frac{t_{\rm bo}^{\rm PL}}{t_{\rm ref}}\right)^{1-\beta}
\right],
\label{eq:Rsh_PL}
\end{equation}

where $R_{\rm bo}^{\rm PL}$ and $t_{\rm bo}^{\rm PL}$ are the shock
radius and time at breakout for the PL parameterization.

Combining the PL velocity evolution with the fitted column density
$N_H(t)$ and substituting them into the general expressions for the
shock power and hadronic interaction efficiency derived in the main
text, the neutrino luminosity can be written as

\begin{equation}
\begin{split}
L_\nu^{\rm PL}(t)
={}&
2\pi \epsilon_{\rm CR}\mu_e m_p
\frac{\alpha A B^2}{t_{\rm ref}}
\left[R_{\rm sh}^{\rm PL}(t)\right]^2
\left(\frac{t}{t_{\rm ref}}\right)^{-\alpha-2\beta-1}
\\
&\times
\left[
1-
\exp\left(
-\frac{\kappa_{pp}\sigma_{pp}\alpha A c}
{B^2 t_{\rm ref}}
R_{\rm sh}^{\rm PL}(t)
\left(\frac{t}{t_{\rm ref}}\right)^{-\alpha+2\beta-1}
\right)
\right].
\end{split}
\label{eq:Lnu_PL}
\end{equation}

Here, $R_{\rm sh}^{\rm PL}(t)$ is given by
Eq.~\ref{eq:Rsh_PL}. Equation~\ref{eq:Lnu_PL} therefore expresses the
neutrino luminosity directly in terms of the parameters describing the
fitted $N_H(t)$ and shock-velocity evolution.

\subsection{Constant+power-law parameterization}

In the const+PL scenario, the shock velocity is assumed to remain
constant at $v_X$ between the SBO epoch and the first X-ray velocity
measurement at $t_X$, and to follow the PL evolution of
Eq.~\ref{eq:v_sh_PL} thereafter. The early time shock radius evolution
is given in the main text by Eq.~\ref{eq:Rsh_constPL_density}.
Extending this evolution beyond $t_X$, the complete shock radius can be
written as

\begin{equation}
R_{\rm sh}^{\rm const+PL}(t)
=
\begin{cases}
\displaystyle
R_{\rm bo}^{\rm const+PL}
+
v_X
\left(t-t_{\rm bo}^{\rm const+PL}\right),
&
t_{\rm bo}^{\rm const+PL}\leq t<t_X,
\\[1.2em]
\displaystyle
R_X^{\rm const+PL}
+
\frac{B t_{\rm ref}}{1-\beta}
\left[
\left(\frac{t}{t_{\rm ref}}\right)^{1-\beta}
-
\left(\frac{t_X}{t_{\rm ref}}\right)^{1-\beta}
\right],
&
t\geq t_X,
\end{cases}
\label{eq:Rsh_constPL}
\end{equation}

where

\begin{equation}
R_X^{\rm const+PL}
=
R_{\rm bo}^{\rm const+PL}
+
v_X
\left(t_X-t_{\rm bo}^{\rm const+PL}\right)
\end{equation}

is the shock radius at the first X-ray velocity measurement.

Combining this shock evolution with the fitted column density $N_H(t)$
and substituting them into the general expressions for the shock power
and hadronic interaction efficiency derived in the main text, the
neutrino luminosity can be written as

\begin{equation}
L_\nu^{\rm const+PL}(t)
=
\begin{cases}
\displaystyle
2\pi \epsilon_{\rm CR}\mu_e m_p
\frac{\alpha A v_X^2}{t_{\rm ref}}
\left[R_{\rm sh}^{\rm const+PL}(t)\right]^2
\left(\frac{t}{t_{\rm ref}}\right)^{-\alpha-1}
\\[0.5em]
\displaystyle\qquad\times
\left[
1-
\exp\left(
-\frac{\kappa_{pp}\sigma_{pp}\alpha A c}
{v_X^2 t_{\rm ref}}
R_{\rm sh}^{\rm const+PL}(t)
\left(\frac{t}{t_{\rm ref}}\right)^{-\alpha-1}
\right)
\right],
&
t_{\rm bo}^{\rm const+PL}\leq t<t_X,
\\[1.5em]
\displaystyle
2\pi \epsilon_{\rm CR}\mu_e m_p
\frac{\alpha A B^2}{t_{\rm ref}}
\left[R_{\rm sh}^{\rm const+PL}(t)\right]^2
\left(\frac{t}{t_{\rm ref}}\right)^{-\alpha-2\beta-1}
\\[0.5em]
\displaystyle\qquad\times
\left[
1-
\exp\left(
-\frac{\kappa_{pp}\sigma_{pp}\alpha A c}
{B^2 t_{\rm ref}}
R_{\rm sh}^{\rm const+PL}(t)
\left(\frac{t}{t_{\rm ref}}\right)^{-\alpha+2\beta-1}
\right)
\right],
&
t\geq t_X .
\end{cases}
\label{eq:Lnu_constPL}
\end{equation}

At $t\geq t_X$, the PL and const+PL parameterizations share the same
shock-velocity evolution and therefore the same local CSM density.
Their shock radii, however, remain different because they retain the
integrated effect of the different velocity histories between SBO and
$t_X$. Consequently, the uncertainty in the unobserved early-time shock
evolution propagates into the predicted neutrino luminosity even after
the two velocity prescriptions converge.

\clearpage

\bibliography{biblio.bib}{}
\bibliographystyle{aasjournal}

\end{document}